\documentstyle[11pt,aaspp4]{article} 
\slugcomment{To appear in Ap.J. Letters} 
\lefthead{Haisch, Lada, \& Lada}  
\righthead{Disk Lifetimes in Cluster Environments: Constraints on
Planet Formation Timescales}

\begin{document}

\title{Disk Frequencies and Lifetimes in Young Clusters}

\author{Karl E. Haisch Jr.\altaffilmark{1,2,3,4} and Elizabeth A. Lada\altaffilmark{3,5}} 
\affil{Dept. of Astronomy, University of Florida, 211 SSRB, Gainesville, FL 32611}

\and

\author{Charles J. Lada\altaffilmark{3}} 
\affil{Smithsonian Astrophysical Observatory, 60 Garden Street, 
Cambridge, Massachusetts 02138}

\altaffiltext{1}{Current address: NASA Ames Research Center, Mail Stop 245-6,
Moffett Field, California 94035-1000}

\altaffiltext{2}{NASA Florida Space Grant Fellow}

\altaffiltext{3}{Visiting Astronomer, Fred Lawrence Whipple Observatory.}

\altaffiltext{4}{Visiting Astronomer at the Infrared Telescope Facility which is
operated by the University of Hawaii under contract to the National Aeronautics
and Space Administration.}

\altaffiltext{5}{Presidential Early Career Award for Scientists and Engineers
Recipient.}

\begin{abstract} We report the results of the first sensitive {\it L}-band
survey of the intermediate age (2.5 -- 30 Myr) clusters NGC 2264, NGC 2362 and NGC
1960.  We use {\it JHKL} colors to obtain a census of the circumstellar disk
fractions in each cluster.  We find disk fractions of 52\% $\pm$ 10\%, 12\%
$\pm$ 4\% and 3\% $\pm$ 3\% for the three clusters respectively.  Together with
our previously published {\it JHKL} investigations of the younger NGC 2024,
Trapezium, and IC 348 clusters, we have completed the first systematic and
homogenous survey for circumstellar disks in a sample of young clusters that both span a significant range in age (0.3 -- 30 Myr) and contain statistically
significant numbers of stars whose masses span nearly the entire stellar mass
spectrum.  Analysis of the combined survey indicates that the cluster disk
fraction is initially very high ($\geq$ 80\%) and rapidly decreases with increasing cluster age, such that half the stars within the clusters lose their disks in $\la$ 3 Myr. Moreover, these observations
yield an overall disk lifetime of $\sim$ 6 Myr in the surveyed cluster sample.
This is the timescale for essentially all the stars in a cluster to lose their
disks. This should set a meaningful constraint for the planet building
timescale in stellar clusters.  The implications of these results for current
theories of planet formation are briefly discussed.  \end{abstract}

 
\section{Introduction}

The fraction of stars presently being formed in the Galaxy that will develop
accompanying
planetary systems, f$_{ps}$, is the product of the fraction of stars initially
surrounded by disks, f$_{i}$, and the fraction of original disks that will
evolve to form planets, f$_{p}$, i.e., f$_{ps}$ = f$_{i}$ $\times$ f$_{p}$.  
Both quantities (i.e., f$_{i}$ and f$_{p}$) may be
functions of stellar mass, among other astrophysical parameters. The
initial disk fraction (f$_{i}$) around newly formed stars can be determined
directly from observations. However, to determine the probability of planet formation
(f$_{p}$) in such disks requires both empirical and theoretical knowledge of the
planet formation process. As a first step, empirical
knowledge of disk evolution can be derived from observations of young stars of
varying age and used to determine disk lifetimes and thus the amount of time
available for planet building around these objects.  Such information should
provide critical tests of planet formation theories and ultimately constraints
for f$_{p}$.

Infrared studies of individual star forming regions such as Ophiuchus, Taurus,
Mon R2, NGC 1333, Orion, IC 348, Sco-Cen and NGC 2362 have long suggested both
that the initial disk frequency is relatively high (f$_{i}$ $>$ 50\%; \cite{wly89}, 
\cite{sta94}, \cite{kh95}, \cite{lal96}, \cite{car97}) and that disk lifetimes are 
relatively short, (3-15 $\times $ 10$^6$ yrs.; \cite{str89}, \cite{sk90}, 
\cite{ses93}, \cite{ll95}, \cite{bran00}, \cite{alv01}).  However, 
these studies suffer from a number of limitations, such as incomplete samples of 
stars, inhomogenous data, uncertain age determinations, etc., which make it 
difficult to draw firm conclusions about either the exact value of the initial disk 
frequency or the lifetime of the circumstellar disk phase of early stellar evolution.  
In particular, most of these studies are based on statistics derived from near-infrared 
{\it JHK} (1.25, 1.65, 2.2 $\mu$m) observations which are not long enough in wavelength 
to produce a complete or unambiguous census of disk bearing stars in a population of 
young stellar objects.  Thus, existing studies provide tantilizing, but not particularly 
strong, constraints to guide planet formation theories.

It is clearly desirable to extend and improve 
upon existing work and place the question of disk evolution on a more firm footing.  
To do so requires a systematic observational campaign which both: 1) is conducted at 
infrared wavelengths which are sufficiently long to identify
disk bearing stars with the minimum of ambiguity and 2) includes a statistically
meaningful sample of stars covering a sufficient range of (relatively well
determined) ages.

\section{The L-band Cluster Survey}

In order to determine a more robust estimate of disk fractions and lifetimes in
star-forming environments, we have embarked on a program to produce a homogenous
and systematic {\it L}-band (3.4 $\mu$m) imaging survey of a sample of nearby,
young, mostly embedded, clusters.  Not only are such clusters believed to be the
birth sites of most galactic field stars (\cite{lada92}, \cite{car00}), but they
also individually and collectively contain statistically significant numbers
(hundreds) of stars, providing a meaningful sampling of essentially the entire
stellar mass function.  Moreover, a survey of clusters enables a more meaningful
determination of the disk evolution timescale than is possible from a survey of
a single star-forming region.  This is because:  1) a cluster can be
characterized by the mean age of all its members, which statistically is a
relatively reliable indicator of age, and 2) a sample of clusters can be 
observed which span a significantly wider range of age than is typical of stars 
formed in any individual star-forming region. Finally, our previous studies have
demonstrated that {\it L}-band observations are capable of detecting and
identifying essentially all the disk bearing stars in a population of young
stellar objects (\cite{hll00}; hereafter HLL00, \cite{la00}, \cite{hai01}). The fraction of stars
with circumstellar disks is given directly by the fraction of stars
with measured {\it L}-band excess emission.

Our survey sample consists of six nearby and relatively rich young clusters 
whose published mean ages span a range between 0.3 -- 30 Myr.
In previous contributions, we have presented the results of {\it L}-band observations of 
the NGC 2024, Trapezium, and IC 348 clusters (mean ages of $\sim$ 0.3 -- 2 Myr) which 
clearly demonstrated that the initial disk frequency is quite high (f$_i $$>$ 80\%) and 
independent of stellar mass, at least in these very young clusters (HLL00, \cite{la00}, 
\cite{hll01}; hereafter HLL01).  These results also indicated that as the age of the 
clusters increased, the disk fraction decreased. The rate of this decrease also 
appeared to be a function of stellar mass, where the disks surrounding high mass stars 
have shorter lifetimes than the disks around stars of lower mass (HLL01).  

In this paper, we present the results of our {\it L}-band imaging surveys of the
older NGC 2264, NGC 2362, and NGC 1960 clusters (mean ages 2.5 -- 30 Myr).  We use
{\it JHKL} colors to obtain a census of the infrared excess/circumstellar disk 
fractions in each
cluster.  In conjunction with our previously published {\it JHKL} data for the
younger NGC 2024, Trapezium, and IC 348 clusters, we are able to 
investigate the evolution of
the cluster disk fraction as a function of age. This study represents the first homogenous and systematic {\it L}-band survey of a 
sample of young clusters which span an adequate age range to enable a straightforward 
determination of the timescale for disk evolution in clusters.

\section{Observations}

We obtained {\it JHK} and {\it L}-band observations for each cluster in our
study.  The {\it L}-band data for each cluster, with the exception of NGC 2362, 
were taken with STELIRCAM (\cite{tw98}) on the Fred Lawrence Whipple Observatory 
(FLWO) 1.2 meter telescope on Mt.  Hopkins, Arizona.  The details of the STELIRCAM 
instrument are discussed elsewhere (HLL00, \cite{la00}). The basic details
of our {\it L}-band image reduction procedures for NGC 2024, the Trapezium, and
IC 348 are described elsewhere (HLL00, Lada et al 2000, HLL01) and, except as 
noted below, are
the same for the NGC 2264 and NGC 1960 clusters. The {\it L}-band data for 
NGC 2362 were obtained using NSFCAM at the NASA Infrared Telescope Facility (IRTF) 
on Mauna Kea, Hawaii (the NSFCAM instrument and image reduction
procedures are discussed in HLL00).

For the NGC 2264 and NGC 1960 clusters, nine fields were observed in a
3$\times$3 square grid which covered an area of $\sim$42 arcmin$^{2}$.  
Note that our observations of NGC 2264 include only a small part of the 
entire region (that centered on the southernmost cluster), while we have 
surveyed most of NGC 1960. 
The total integration time for each field in these clusters is 18 minutes. In NGC 2362, twenty-five fields arranged in a 5$\times$5 square covering an area
of $\sim$28 arcmin$^{2}$ were observed (covering most of the cluster).  
The total integration time for each field was $\sim$ 4.5 minutes. Within the respective survey regions of each cluster, we detected 56, 45 and 93 sources at {\it L} band to the completeness limits of our surveys ({\it L} = 12.0 in NGC 2264 and NGC 1960; {\it L} = 14.0 in NGC 2362). We find a field star contamination within similar areas to be 6, 8 and 18 stars in NGC 2264, NGC 1960 and NGC 2362 respectively.

The {\it JHK} observations of the clusters in our sample were obtained at 
various observatories. The NGC 2264 and Trapezium observations were obtained 
using STELIRCAM on the 1.2 meter telescope at FLWO and are discussed in 
Muench et al. (2001) and Lada et al. (2000). {\it J, H} and {\it K$_{s}$} 
(2.1 $\mu$m) observations of the NGC 1960 cluster were taken from the 2MASS 
data archive (http://pegasus.astro.umass.edu). {\it JHK} observations of the 
NGC 2362 cluster were obtained at the European Southern Observatory (ESO) 2.3
 meter telescope and are discussed in Alves et al. (2001). The {\it JHK} data 
 for the NGC 2024,
and IC 348 clusters were obtained with the NOAO 1.3 meter telescope and
are discussed in HLL00, and Lada \& Lada (1995).  

\section{Results}

We have used {\it JHKL} color-color diagrams to derive the corresponding background 
corrected infrared excess fractions for the NGC 2264, NGC 2362 and NGC 1960 clusters.  
An extensive discussion of the method we use to determine
cluster infrared excess fractions from color-color diagrams is given in HLL00.
We find {\it L}-band excess fractions of 52 $\pm$ 10\%, 12 $\pm$ 4\%, and 3 $\pm$ 3\% for the three clusters respectively. These fractions are
significantly lower than those previously derived for the younger Trapezium
and NGC 2024 clusters (HLL00, \cite{la00}) using similar techniques. 
In Table~\ref{fractable}, we list the derived infrared excess fractions
for the entire sample of clusters which we have surveyed in {\it JHKL} bands. 
Also listed in the table is the estimated stellar mass at the respective completeness 
limits in each cluster as well as the mean age of each cluster taken from the 
literature with the corresponding reference. The ages of the Trapezium, IC 348 and NGC 2264 were obtained using 
PMS tracks of Palla \& Stahler (2000). The mean age for NGC 2024 was adopted from Meyer (1996) and was obtained using the PMS tracks of D'Antona \& Mazzitelli (1994). On the other hand, the ages of 
NGC 2362 and NGC 1960 were determined from post-main sequence isochrone 
fitting in the HR diagram.

In Figure~\ref{diskfracs} we plot the {\it JHKL} excess (disk) fraction 
as a function of mean age for all the clusters in our survey except NGC 2362, where we plot the age based on post-main sequence isochrone fitting as discussed earlier. For comparison 
we also plot excess fractions in Taurus and Chamaeleon I (open triangles),  
derived  from similar {\it JHKL} observations in the literature 
(i.e., \cite{kh95}, \cite{kg01}). 
The ages for Taurus and Chamaeleon I were obtained from Palla \& Stahler (2000). NGC 1960 is not included 
in the figure since our observations of this cluster only extend 
to $>$ 1 M$_{\odot}$ stars, whereas in the other clusters we are 
complete to $\leq$ 1.0 M$_{\odot}$.

The dot-dashed line in Figure 1 represents a least squares fit to 
the data obtained in our {\it L}-band survey (filled triangles). Vertical error bars 
represent the statistical $\sqrt{{\it N}}$ errors in our derived 
excess/disk fractions. Horizontal error bars show representative 
errors of our adopted ages. The error bars for the ages of the Trapezium, Taurus, IC 348, Chamaeleon I and NGC 2264 represent the error in the mean of the individual source ages derived from a single set of PMS tracks. In order to estimate the overall systematic uncertainty introduced in using different PMS tracks, we calculated the mean age and the standard deviation of the mean age for NGC 2264 (2.6 $\pm$ 1.2 Myr) from five different PMS models (\cite{psbk00}, \cite{ps00}). This latter quantity illustrates the likely systematic uncertainty introduced by the overall uncertainties in the PMS models. This is plotted in Figure 1. For stars with M $\leq$ 1 M$_{\odot}$ and ages $\leq$ 5 Myr, the overall uncertainty in the ages for all regions is likely within about 1 -- 1.2 Myr. The plotted error for NGC 2024 reflects this uncertainty. The age error for NGC 2362 was adopted from the literature (\cite{bl96}).

\section{Discussion}

We have completed the first sensitive {\it L}-band survey of a sample of 
young clusters that span a sufficent range in age (0.3 -- 30 Myr) to 
enable a meaningful determination of the timescale for disk evolution 
within them. Clusters appear to be characterized by a very high initial 
disk frequency ($\geq$ 80\%) which then sharply decreases with cluster age. 
Half the disks in a cluster population are lost in only about 3 Myr, and the 
timescale for essentially all the stars to lose their disks appears to be about 6 Myrs. 

The precise value of this latter timescale to some extent depends on the derived 
parameters for the NGC 2362 cluster. Our quoted timescale of 6 Myrs could be 
somewhat of a lower limit for two reasons. First, it is possible that a 
slightly higher disk fraction for NGC 2362 could be obtained with deeper 
{\it L}-band observations which better sample the cluster population below 
1 M$_{\odot}$. Our earlier observations of IC 348 and the Trapezium cluster 
show that the disk lifetime appears to be a function of stellar mass 
(HLL01), with higher mass stars losing their disks faster than lower mass 
stars. However, we note that much deeper {\it JHK} observations (\cite{alv01}) 
which sample the cluster membership down to the 
hydrogen burning limit yield a {\it JHK} disk fraction of essentially
0\%, giving us confidence in the very low disk fraction derived from our 
present {\it L}-band observations. Second, the age of NGC 2362 is dependent 
on the turnoff age assigned to only one star, the O star, $\tau$ CMa. 
This star is a multiple system and its luminosity assignment on the HR diagram 
somewhat uncertain (\cite{vlg97}). Correction for multiplicity would lead to a 
slightly older age. However, the quoted 1 Myr error in its age likely reflects the magnitude of this uncertainty (\cite{bl96}). On the other hand, if, for example, the errors were twice as large as quoted, the cluster could have an age between 3 Myr and 7 Myr. The corresponding age and the overall disk lifetime derived from a least squares fit to the data would be between 4 Myr and 8 Myr. Even if the timescale for all disks to be lost was as large as 8 Myr, our survey data would still require that half the stars lose their disks on a timescale $<$ 4 Myr. Finally, an even older age for NGC 2362 would likely indicate that the decrease in disk fraction with time does not follow a single linear fit, that is, after a rapid decline during which most stars lost their disks, the disk fraction in clusters would decrease more slowly, with a small number of stars ($\sim$ 10\%) retaining their disks for times comparable to the cluster age. On the other hand, we estimate the dynamical age of the S310 HII region which surrounds, and is excited by, $\tau$ CMa to be $\sim$ r$_{HII}$ / v$_{exp}$ $\sim$ 5 $\times$ 10$^{6}$ yrs for r$_{HII}$ = 50 pc and v$_{exp}$ = 10 km s$^{-1}$ (e.g. \cite{lr78}, \cite{jbn98}). This is consistent with the turn-off age of the cluster derived from the HR diagram, and supports our estimate of $\sim$ 6 Myr for the overall disk lifetime.

We point out that our {\it L}-band observations directly measure 
the excess caused by the presence of small (micron-sized), 
hot ($\sim$ 900 K) dust grains in 
the inner regions of the circumstellar disk and these observations are sensitive
to very small amounts ($\sim$ 10$^{20}$ gm) of dust. We expect that if there 
is gas in the disk, turbulent motions  will always keep significant
amounts of small dust particles mixed with the gas (\cite{rud99}), thus dust 
should remain a good tracer of gas in the disks as they evolve to form planets. Indeed, 
recent observations of H$_2$ in older debris disks appear to confirm this
assertion (\cite{thi01}). Consequently, stars which did 
not show infrared excesses are likely to be significantly devoid of gas as 
well as dust. There is also evidence that the presence of dust in the inner 
disk regions is linked with the presence of dust in the outer disk 
regions (i.e., r $>$ 1 AU) where most planet formation is likely to occur. 
Earlier, HLL01 noted a strong correlation between the presence of 
an {\it L}-band excess  and millimeter-wave 
continuum emission (which traces the outer disk regions) for stars in the 
Taurus clouds. Thus the disk evolution timescales derived from 
our {\it L}-band survey likely characterize the bulk of the disk material
around a young star and, consequently, should place important constraints on the 
timescale allowed for building gas giant planets around such stars.

For example, according to the accepted model of giant planet formation, the core
accretion model, such planets are built in the following sequence: 
a high density (rocky) core with a mass of $\sim$ 10 Earth masses forms, followed by the accumulation of substantial 
amounts of nebular gas.  
Typical timescales derived from these models range from a few million years 
to greater than 10 Myr depending on the mass accretion rate of solid material, 
the disk surface density of solid material and atmospheric dust 
abundance (\cite{bhl00}, \cite{liss01}).  A disk lifetime of $\sim$ 6 Myr 
sets strong constraints on the range of possible values for 
these parameters.  The constraints are even more stringent for the highest 
mass stars, since these stars may lose their disks on even shorter 
timescales ($<$ 2 -- 3 Myr; HLL01). If there is sufficient gas remaining in the 
disk after the planet forms, the planet may then migrate inward toward the central 
star (\cite{lbr96}). The discovery of numerous extrasolar planets in very
close proximity to their central stars (e.g., \cite{mar99})
would therefore seem to require an even more extended disk lifetime. 
It may be very difficult to both build 
and then migrate planets on such short timescales with standard core accretion models.
A competing model can produce giant planets on a 
much shorter ($\sim$ 10$^{3}$ yr) timescale via gravitational 
instability within a relatively massive (M$_{disk}$ $\sim$ 0.1 M$_{\odot}$ 
within a radius of 20 AU) protoplanetary disk (Boss 1998, 2000). 
The disk lifetimes we measure would be sufficient for giant planets to form
via gravitational instability. However, whether such models are really
viable requires more theoretical investigation (\cite{bos00}). If not, forming 
giant planets in disks may be more difficult than currently envisioned, and the
probabilty (f$_p$) of producing giant planets in typical circumstellar disks
may be relatively small. 

However, we cannot firmly rule out the possibility that the lifetime for 
the outer disk regions, which contain the bulk of the mass available to form planets, 
may be different from that for the inner disk. If the outer disk lifetime is
longer than that which we measure for the inner disk, models which predict a longer giant planet
formation timescale could be accomodated (\cite{thi01}, \cite{liss01}).  Longer
wavelength millimeter and Space Infrared Telescope Facility (SIRTF) observations
of statistically significant samples of YSOs in a number of young clusters 
would be very beneficial to providing further insight into this question and
improving the assessment of the overall likelihood of planet formation in 
circumstellar disks.
 
\acknowledgements

We thank August Muench for obtaining the {\it JHK} observations of the NGC 2264
cluster and Jo\~{a}o Alves for providing near-infrared
{\it JHK} photometry of the NGC 2362 cluster in advance of publication. We also thank Richard Elston for many useful discussions. K. E. H. gratefully acknowledges support from a NASA Florida Space Grant Fellowship
and an ISO grant through JPL \#961604. E. A. L. acknowledges support from a
Research Corporation Innovation Award and a Presidential Early Career Award for
Scientists and Engineers (NSF AST 9733367) to the University of Florida.  We
also acknowledge support from an ADP (WIRE) grant NAG 5-6751.   

\newpage

\begin{deluxetable}{cccl}
\tablecaption{Background Corrected {\it JHKL} Cluster Infrared Excess Fractions \label{fractable}}
\tablewidth{0pt}
\tablehead{Cluster & {\it JHKL} Excess Fraction\tablenotemark{1} & Mass (M$_{\odot}$)\tablenotemark{2} & Age (Myr) \& Reference\tablenotemark{3}}
\startdata
NGC 2024 & 112/131 (85\% $\pm$ 8\%) & 0.13 & 0.3; \cite{mey96}\nl
Trapezium & 312/391 (80\% $\pm$ 5\%) & 0.16 & 1.5; \cite{ps00}\nl
IC 348 & 59/91 (65\% $\pm$ 8\%) & 0.19 & 2.3; \cite{ps00}\nl
NGC 2264 & 26/50 (52\% $\pm$ 10\%) & 0.85 & 3.2; \cite{ps00}\nl
NGC 2362 & 9/75 (12\% $\pm$ 4\%) & 1.00 & 5.0; \cite{bl96}\nl
NGC 1960 & 1/37 (3\% $\pm$ 3\%)\tablenotemark{4} & 1.30 & 30.0; \cite{bzso85}\nl
\enddata
\tablenotetext{1}{A {\it K$_{s}$} filter was used for the observations of NGC 1960.}
\tablenotetext{2}{Mass of a star at the completeness limit for each cluster.}
\tablenotetext{3}{Reference from which cluster age was taken.}
\tablenotetext{4}{The source which showed an {\it L}-band excess is an outflow source superimposed on the cluster field and likely not associated with the cluster itself (\cite{mag96}).}
\end{deluxetable}

\newpage 

\figcaption[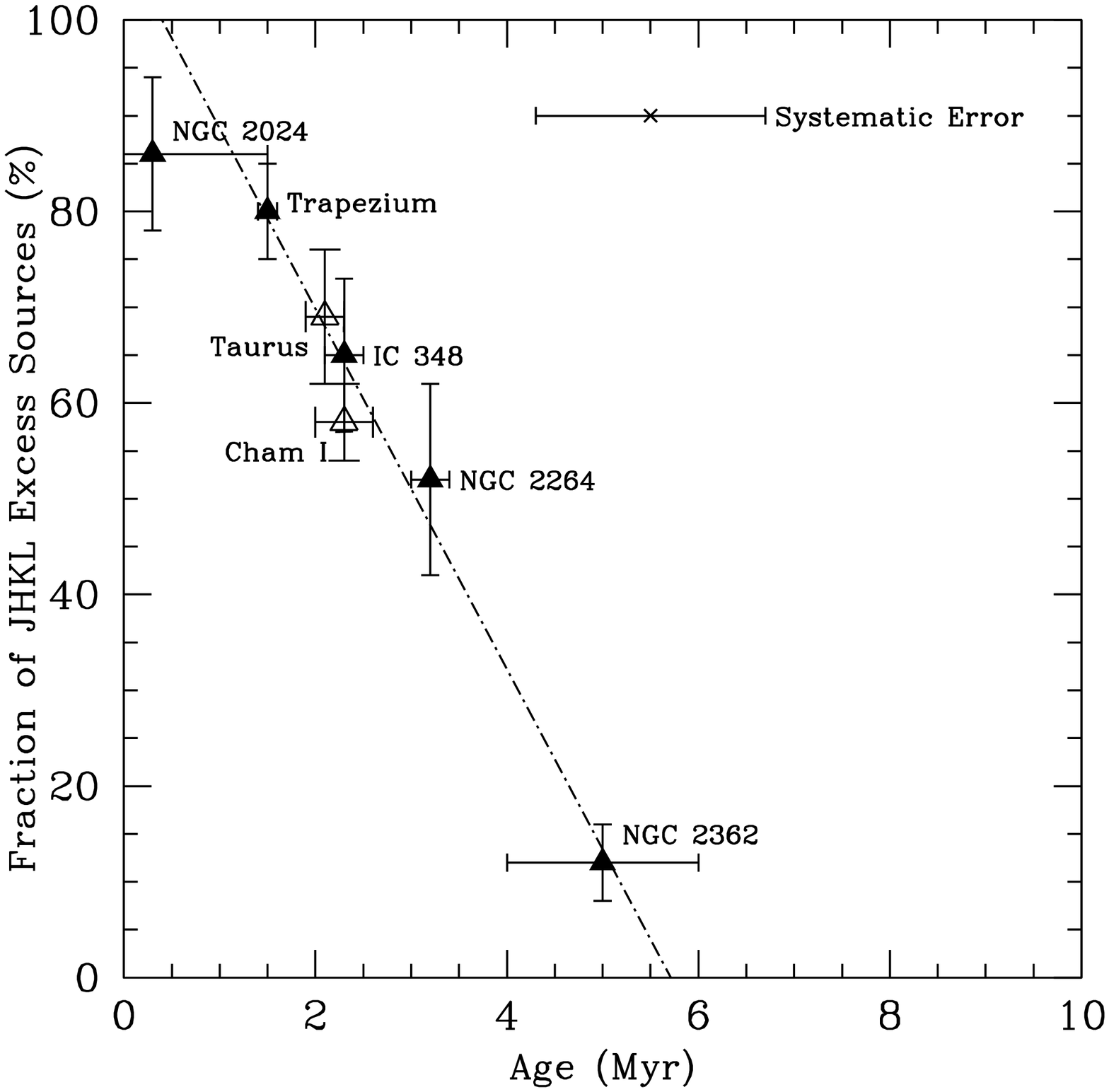]
{The {\it JHKL} excess/disk fraction as a function of mean cluster age. Vertical error bars represent the statistical $\sqrt{{\it N}}$ errors in our derived 
excess/disk fractions. Horizontal error bars show representative errors of our adopted ages. See text for discussion.
\label{diskfracs}
}
\clearpage
\plotone{Haisch.fig1.eps}
\clearpage

\end{document}